%% file: hops_phot_arxiv.tex
\documentclass[twocolumn]{aastex61}
\pdfoutput=1

\shorttitle{Protostellar Properties across Orion}
\shortauthors{Fischer et al.}

\begin{document}

\newcommand{\Herschel}{\textit{Herschel}}
\newcommand{\Spitzer}{\textit{Spitzer}}

\title{The \textit{Herschel}\footnote{\Herschel\ is an ESA space observatory with science instruments provided by European-led Principal Investigator consortia and with important participation from NASA.} Orion Protostar Survey: Far-Infrared Photometry and Colors\\ of Protostars and Their Variations across Orion A and B}

\correspondingauthor{William J. Fischer}
\email{wfischer@stsci.edu}

\author[0000-0002-3747-2496]{William J. Fischer}
\affil{Space Telescope Science Institute, Baltimore, MD, USA}

\author[0000-0001-7629-3573]{S. Thomas Megeath}
\affil{Ritter Astrophysical Research Center, Department of Physics and Astronomy, University of Toledo, Toledo, OH, USA}

\author[0000-0001-9800-6248]{E. Furlan}
\affil{NASA Exoplanet Science Institute, Caltech/IPAC, Pasadena, CA, USA}

\author[0000-0003-2300-8200]{Amelia M. Stutz}
\affil{Max-Planck-Institut f\"ur Astronomie, Heidelberg, Germany}
\affil{Departmento de Astronom\'ia, Facultad de Ciencias F\'isicas y Matem\'aticas, Universidad de Concepci\'on, Concepci\'on, Chile}

\author[0000-0002-5812-9232]{Thomas Stanke}
\affil{European Southern Observatory, Garching bei M\"unchen, Germany}

\author[0000-0002-6195-0152]{John J. Tobin}
\affil{National Radio Astronomy Observatory, Charlottesville, VA, USA}

\author[0000-0002-6737-5267]{Mayra Osorio}
\affil{Instituto de Astrof\'isica de Andaluc\'ia, CSIC, Granada, Spain}

\author[0000-0002-3530-304X]{P. Manoj}
\affil{Department of Astronomy and Astrophysics, Tata Institute of Fundamental Research, Mumbai, India}

\author[0000-0002-9289-2450]{James Di Francesco}
\affil{Department of Physics and Astronomy, University of Victoria, Victoria, BC, Canada}
\affil{NRC Herzberg Astronomy and Astrophysics, Victoria, BC, Canada}

\author[0000-0002-7789-5119]{Lori E. Allen}
\affil{NSF's NOIRLab, Tucson, AZ, USA}

\author[0000-0001-8302-0530]{Dan M. Watson}
\affil{Department of Physics and Astronomy, University of Rochester, Rochester, NY, USA}

\author[0000-0003-1755-8759]{T. L. Wilson}
\affil{Max-Planck-Institut f\"ur Radioastronomie, Bonn, Germany}

\author[0000-0002-1493-300X]{Thomas Henning}
\affil{Max-Planck-Institut f\"ur Astronomie, Heidelberg, Germany}

\begin{abstract}
The degree to which the properties of protostars are affected by environment remains an open question. To investigate this, we look at the Orion A and B molecular clouds, home to most of the protostars within 500 pc. At $\sim$ 400 pc, Orion is close enough to distinguish individual protostars across a range of environments in terms of both the stellar and gas projected densities. As part of the \Herschel\ Orion Protostar Survey (HOPS), we used the Photodetector Array Camera and Spectrometer (PACS) to map 108 partially overlapping square fields with edge lengths of 5$\arcmin$ or 8$\arcmin$ and measure the 70 \micron\ and 160 \micron\ flux densities of 338 protostars within them. In this paper we examine how these flux densities and their ratio depend on evolutionary state and environment within the Orion complex. We show that Class 0 protostars occupy a region of the 70 \micron\ flux density versus 160 \micron\ to 70 \micron\ flux density ratio diagram that is distinct from their more evolved counterparts. We then present evidence that the Integral-Shaped Filament (ISF) and Orion B contain protostars with more massive envelopes than those in the more sparsely populated LDN 1641 region. This can be interpreted as evidence for increasing star formation rates in the ISF and Orion B or as a tendency for more massive envelopes to be inherited from denser birth environments. We also provide technical details about the map-making and photometric procedures used in the HOPS program.\end{abstract}


\section{Introduction}

A complete picture of star formation, from the gravitational collapse of a molecular cloud to the dispersal of the circumstellar envelope and disk, is a fundamental part of our understanding of our cosmic origins. In recent years, space-based mid- to far-infrared (far-IR) surveys have mapped a large sample of protostars in nearby molecular clouds, enabling detailed studies of protostellar evolution \citep{dun14}. In this protostellar phase, a dusty infalling envelope and a nascent protoplanetary disk surround an accreting protostar. The envelope absorbs short-wavelength radiation from the central protostar and reprocesses most of the luminosity to far-IR wavelengths \citep[e.g.,][]{whi03}. The protostar drives a bipolar outflow that may evacuate cavities in the envelope, allowing some of the shorter wavelength radiation to escape. After $\sim$ 0.5 Myr, the envelope disappears and the protostellar phase is over. It is not clear to what extent the dissipation of the envelopes is driven by feedback from outflows or the depletion of the reservoir of gas in the environment. 

Far-IR observations are an essential tool for understanding the protostellar phase. Protostars typically radiate most of their luminosity in the far IR, the dusty star-forming environment also emits strongly at far-IR wavelengths, and far-IR photons easily penetrate this dust to reach the observer. During the protostellar stage, the forming star is still strongly connected to the local environment, and variations in the gas environment may alter the trajectory of protostellar evolution. A detailed characterization of protostellar evolution in the far IR thus allows us to better understand the role of environment in star formation.

This link between protostars and their surroundings presents challenges to observational studies. Observers must disentangle environmental emission from that of the protostar. The spatial scales of the central object, disk, envelope, and outflows range from less than one AU to thousands of AU. Observational strategies must account for morphologies that are complex at both compact and extended scales.

The \Herschel\ Orion Protostar Survey (HOPS; PI: S. T. Megeath) was a 200 hour open-time key program of the \Herschel\ Space Observatory \citep{pil10}. HOPS used the Photodetector Array Camera and Spectrometer (PACS) instrument \citep{pog10} to obtain 70 and 160 \micron\ images (with angular resolutions of 5.2\arcsec\ and 12\arcsec, respectively) and 55--200 \micron\ spectra of protostars identified in the \Spitzer\ Space Telescope survey of Orion \citep{meg12,meg16}.

The use of \Herschel\ data facilitates the direct measurement of emission from the protostellar envelopes, sampling the peaks of their spectral energy distributions (SEDs). The unprecedented sensitivity afforded by \Herschel\ at far-IR wavelengths allowed us to efficiently survey a large number of protostars \citep{fur16} and discover new ones \citep{stu13}. We supplemented these data with imaging, photometry, and spectroscopy from 1.2 to 870~\micron. Multiwavelength data allow us to constrain the protostars and the properties of their envelopes \citep{fur16}.

A major benefit to the study of Orion is that it contains a large sample of protostars in a single cloud complex. \citet{fur16} tabulated 319 protostars (in a sample of 330 young stellar objects or YSOs) in Orion alone, while \citet{dun13} used \Spitzer\ data to find 230 protostars in 18 other nearby ($<$ 500 pc) molecular clouds.

A second benefit to the study of Orion is that, at approximately 400 pc, it is relatively nearby. According to \citet{kou18}, our targets mostly lie at distances from 389 pc to 417 pc, except for a few at 345~pc; see further discussion below. \Herschel\ data therefore provide sufficiently high spatial resolution to isolate individual protostellar envelopes in a single comparatively high-mass ($\sim$~10$^5$ $M_\sun$; e.g., \citealt{stu15}) and nearby cloud, even in clustered regions. The only similarly nearby and massive cloud is the California molecular cloud \citep{lad09}, but it contains far fewer YSOs. At the distance of Orion, the angular resolution of PACS corresponds to distances of 2100 and 4800 AU for the 70 and 160 \micron\ channels, respectively.

A third benefit to the study of Orion is that it contains significantly different environments, from isolated cold globules to rich clusters. There are two common observational measures of environment. First, there are the environmental conditions set by the properties of the natal molecular gas, particularly the dense gas. Second, there are environmental variations traced by the protostars, including the densities of young stars and the systematic changes in their properties.  These protostars and their properties have been quantified with \Spitzer\ data \citep{meg12,meg16} and \Herschel\ data from HOPS \citep{fur16}.

With \Herschel\ observations at longer wavelengths and of wider fields than included in HOPS, \citet{stu15} measured the column density ($N_{\rm H}$) and mass distributions across Orion A, quantifying environmental differences. They found that the column density probability distribution functions vary with location. Furthermore, \citet{stu16} and \citet{stu18} found that the mass per unit length ($M/L$) varies significantly across Orion A, with the Orion Nebula Cluster (ONC) having a higher $M/L$ than the Integral Shaped Filament (ISF), and the ISF having a higher $M/L$ than LDN 1641. These column density and mass variations imply variations in the volume density and gravitational potential across the cloud.

Environmental variations are also found in the radial velocities of the gas. \citet{gon19} showed that gas radial velocities in both high- and low-density tracers show significant variations within the ISF. In particular, the northern portion has more centrally concentrated gas with stronger overall velocity gradients compared to the southern portion, which transitions to the LDN 1641 region. Systematic variations of the temperatures, line widths, and densities of cores across Orion A were also found by \citet{wil99}.

Turning to the stellar content, the observed properties of the protostars vary with environment. \citet{kry12} found variations in the Orion clouds, where protostars are more luminous in dense regions. \citet{stu15} found that the fraction of protostars in the young Class~0 phase varies systematically with changes in environment as indicated by variations in the slope of the dust column density probability distribution function ($N$-PDF) across the region. The variation in this fraction may result from differences in the star formation history between these regions, as also suggested by \citet{fis17}, the influence of feedback \citep{sad14}, or variations in infall rates of the protostars due to variations in the densities of their birth environments \citep{kry12,kry14,dun14}.

This paper is part of a series describing results from HOPS. These papers include evolutionary studies of protostars via modeling of their SEDs \citep{fur16,fis17}, the analysis of far-IR protostellar spectra \citep{man13,man16}, the discovery and characterization of the youngest protostars \citep{stu13,tob15,tob16,kar20}, a study of multiplicity in the Orion YSOs \citep{kou16}, progress on understanding outburst phenomena in protostars \citep{fis12,saf15,fis19}, and detailed studies of the active OMC 2/3 region \citep{fur14,gon16,oso17}.

Here we present the 70 and 160 \micron\ photometry obtained with \Herschel\ and first reported by \citet{fur16} as part of their effort to model the SEDs of the Orion protostars with \Herschel\ and other multiwavelength data. Section~\ref{s.sample} reviews the main aspects of the sample selection and observing strategy, pointing the reader to the Appendix for the HOPS catalog and previously unpublished descriptions of the map-making and photometric techniques that lie at the foundation of \citet{fur16} and other HOPS papers.

Section~\ref{s.trends} uses the HOPS 70 and 160 \micron\ photometry to analyze protostellar properties across the Orion complex, including their dependence on evolutionary stage and location within Orion. In part, it relies on the SED classification by \citet{fur16}. Section~\ref{s.disc} contains our discussion, and our conclusions are summarized in Section~\ref{s.conc}. Our maps, photometry, SEDs, and model fits to the SEDs can be found at the Infrared Science Archive (IRSA).\footnote{\url{https://irsa.ipac.caltech.edu/data/Herschel/HOPS/overview.html}}

\section{Sample Selection}\label{s.sample}

The HOPS sources are numbered from 0 to 409. Most were identified as protostars in the \Spitzer\ survey of Orion. The \Spitzer\ protostar sample was defined and described by \citet{meg12}, with minor modification by \citet{meg16}. \citet{kry12} discussed the development of the initial criteria. Their identification relied on \Spitzer\ 3.6--24~\micron\ photometry merged with 1--2~\micron\ photometry from the Two Micron All Sky Survey (2MASS; \citealt{skr06}) point-source catalog. Sixteen targets, in contrast, are \Herschel-identified protostars that showed weak or no detections at wavelengths $\le$~24~\micron\ but were found to be bright in the PACS 70~\micron\ band \citep{stu13,tob15}.

Based on the time awarded for the key program, the \Herschel\ observations were designed to observe protostars from the \Spitzer\ compilation with estimated 70~\micron\ flux densities greater than 42 mJy. Of the 410 numbered sources, 373 were observed. Of these, 337 were detected at 70 \micron\ and 254 were detected at 160 \micron. \citet{fur16} discuss in greater detail the likely nature of sources that were not observed or were observed but not detected. For their study, they focused on the 330 HOPS targets among those detected at 70~\micron\ that they classified as YSOs, 319 of which were determined to be Class~0, Class~I, or flat-spectrum protostars based on their mid-IR spectral indices and bolometric temperatures, and 11 of which were determined to be Class II objects. The other seven 70 \micron\ detections were judged to be extragalactic contaminants (six) or of uncertain nature (one).

\begin{figure}
\includegraphics[width=\hsize]{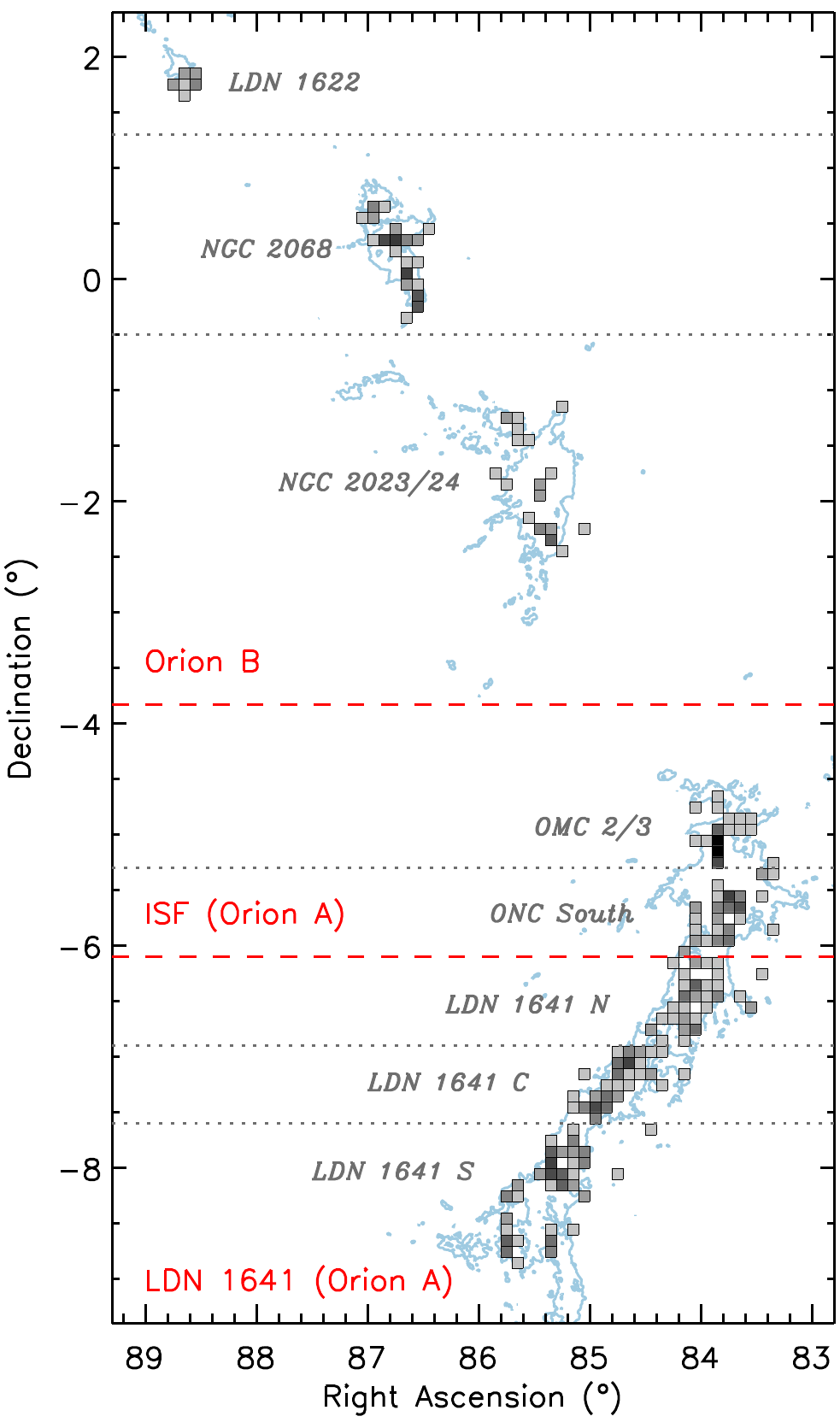}
\caption{Locations within Orion of the 410 HOPS sources. Each box covers $0.1^\circ \times 0.1^\circ$ and is shaded according to the number of HOPS sources within its bounds. The lightest boxes contain single HOPS sources, while the darkest box contains 21. Gray labels and dotted lines indicate the regions listed in Table~\ref{t.regions}, while red labels and dashed lines show how they are combined into super-regions in Section~\ref{s.loc}. The 1.5 Jy beam$^{-1}$ contours from the \Herschel\ 500 \micron\ map are shown in blue \citep{stu15}.\label{f.regions}}
\end{figure}

Table~\ref{t.regions} divides the Orion A and B clouds into regions based on their declinations and shows the number of HOPS targets in each region, the number observed, and the numbers detected at 70 \micron\ and 160 \micron. It additionally shows how many of the 70 \micron\ detections are classified as Class 0, Class~I, or flat-spectrum protostars by \citet{fur16}. Figure~\ref{f.regions} shows how the 410 HOPS sources are distributed across Orion and divided into regions.

\begin{deluxetable*}{@{\extracolsep{4pt}}lcccccccc@{}}
\tablewidth{0pt}
\tablecaption{Regions in Orion and Source Counts\label{t.regions}}
\tablehead{
\colhead{} & \colhead{} & \colhead{Numbered} & \colhead{Observed} & \multicolumn{2}{c}{Detections} & \multicolumn{3}{c}{Protostars\tablenotemark{2}} \\
\cline{5-6}
\cline{7-9}
\colhead{Region} & \colhead{Declination Range} & \colhead{Targets\tablenotemark{1}} & \colhead{Targets} & \colhead{70 \micron} & \colhead{160 \micron} & \colhead{Class 0} & \colhead{Class I} & \colhead{Flat-Spectrum}
}
\startdata
\multicolumn{9}{c}{Orion B} \\
\hline
LDN 1622\tablenotemark{3}    & ($+$1.3, $+$2.1) & 11 & 10 & 9 & 9 & 2 & 6 & 1 \\
NGC 2068    & ($-$0.5, $+$1.3) & 59 & 56 & 53 & 45 & 22 & 21 &  8 \\
NGC 2023/24 & ($-$3.83, $-$0.5) & 27 & 20 & 19 & 14 &  8 &  5 &  5 \\
\hline
\multicolumn{9}{c}{Orion A} \\
\hline
OMC 2/3     & ($-$5.3, $-$3.83) & 65 & 60 & 47 & 28 & 16 & 12 & 16 \\
ONC South   & ($-$6.1, $-$5.3) & 54 & 49 & 35 & \phantom{$^4$}24\tablenotemark{4} & 11 & 12 &  9 \\
LDN 1641 N  & ($-$6.9, $-$6.1) & 47 & 42 & 41 & 29 & 10 & 16 & 13 \\
LDN 1641 C  & ($-$7.6, $-$6.9) & 62 & 58 & 55 & 41 & 14 & 21 & 18 \\
LDN 1641 S  & ($-$9.0, $-$7.6) & 85 & 78 & 78 & 64 &  9 & 32 & 32 \\
\hline
Total & ($-$9.0, $+$2.1) & 410 & 373 & 337 & 254 & 92 & 125 & 102 \\
\enddata
\tablenotetext{1}{Four targets are duplicates of other HOPS sources in which nearby scattered light detected with \Spitzer/IRAC was erroneously treated as a unique point source. They are included in this column but no others.}
\tablenotetext{2}{Classification by \citet{fur16} based on mid-IR spectral indices and bolometric temperatures.}
\tablenotetext{3}{Based on spectroscopic and astrometric data, \citet{kou18} conclude that LDN 1622 is in the Orion complex but is not part of Orion B. We include it here for consistency with previous HOPS work; in any case, the number of protostars is small, and this region does not have a major impact on our conclusions.}
\tablenotetext{4}{All 160 \micron\ detections are also 70 \micron\ detections except for one source in ONC South.}
\end{deluxetable*}

\subsection{Mapping Procedure}\label{s.map}

The HOPS targets were divided into distinct spatial groups to optimize observing and were imaged in a series of partially overlapping square maps, either 5\arcmin\ or 8\arcmin\ on a side. The map centers and sizes can be found in Tables 1 and 2 of \citet{stu13}. We used PACS and its scan-map astronomical observing template, with the slowest allowed scan speed (20\arcsec\ s$^{-1}$), to simultaneously obtain 70~\micron\ and 160~\micron\ images. To avoid the striping characteristic of bolometer arrays \citep{teg97}, we mapped each group in two orthogonal scan directions that have consecutive observation identifiers (ObsIDs). Multiple scan legs were needed to cover each group, since they are larger than the 1.75\arcmin\ $\times$ 3.5\arcmin\ PACS field of view. Our first imaging data were obtained for a single field on 2009 October 9 in the science demonstration phase \citep{fis10,sta10}, and subsequent imaging data were obtained between 2010 March 10 and 2011 September~19. In the Appendix, we present the HOPS catalog and discuss the data processing, map generation, and photometric techniques.

\section{Dependence of Photometry on Evolutionary Stage and Environment}\label{s.trends}

Here we examine trends in the 70 \micron\ flux densities and the ratios of 160 \micron\ to 70 \micron\ flux densities of the HOPS protostars as functions of evolutionary stage and region. We use the evolutionary classes assigned by \citet{fur16} based on mid-IR spectral indices and bolometric temperatures. These follow the standard scheme reviewed by \citet{dun14}, where protostars are classified as Class~0, Class I, or flat-spectrum. These classes approximately represent a sequence of evolutionary stages. In the Class~0 protostars, most of the mass is still expected to be in the envelope instead of the star. In contrast, the Class I protostars still have a significant envelope, but most of the mass is expected to be in the star. The flat-spectrum protostars have residual envelopes that typically exceed the masses of their disks. Class II sources are considered to be post-protostellar, although some have residual envelopes.

Since the classification is based on the SED, which may be affected by foreground reddening and the inclination of the protostar, each evolutionary stage may include objects from previous or subsequent classes. \citet{fur16} used the 4.5--24 \micron\ spectral index, which is expected to be relatively unaffected by foreground extinction compared to indices at shorter wavelengths, but they did not explicitly account for extinction in their classification. Although extinction may affect the sorting into  Class II YSOs, flat-spectrum protostars, and Class I protostars, the Class 0 protostars were identified by their low $T_{\rm bol}$. As \citet{stu15} showed statistically, based on the HOPS grid of SED models \citep{fur16}, the rate at which Class~I protostars are misclassified as Class 0 protostars due to foreground extinction is small.

Direct comparison of flux densities is valid if all sources are roughly at the same distance. \citet{kou18} find that the distances to the protostars in our study range from 417 pc at the southern end of LDN~1641 to 389 pc in the vicinity of the ONC, with NGC 2023/24 and NGC 2068 at 403 pc and 417 pc, respectively. They conclude that LDN 1622, at 345 pc, is not part of Orion B; see footnote 2 to Table~\ref{t.regions}. The ratio of the largest to the smallest distance, squared, is 0.06 in logarithmic units, or 0.16 if LDN 1622 is included. This is negligible compared to the range of flux densities considered in this paper. Including more distant Gaia stars in their analyses, subsequent researchers \citep{gro18,zuc20,rez20} reported a distance of 450 pc to the southern end of LDN~1641. This would increase the ratio reported above from 0.16 to 0.23, still small compared to the differences in flux density among regions.

\subsection{Trends with Evolutionary Stage}\label{s.evo}

\begin{figure}
\includegraphics[width=\hsize]{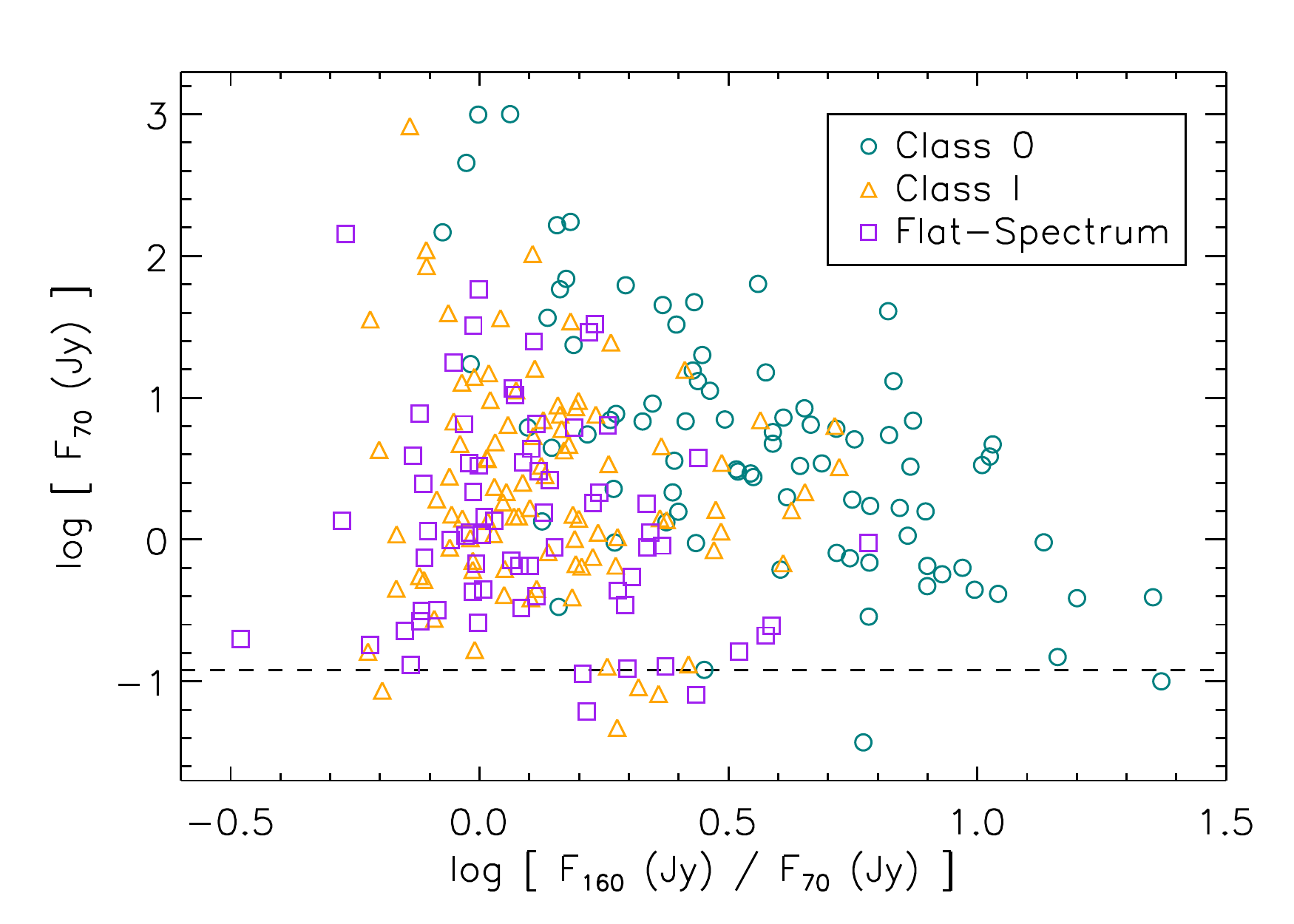}
\caption{Plot of 70 \micron\ flux density against 160 \micron\ to 70~\micron\ flux density ratio for protostars in our sample that were detected at both wavelengths. This includes 84 of 92 Class~0 protostars, 94 of 125 Class~I protostars, and 69 of 102 flat-spectrum protostars. The dashed line at $F_{70}=0.12$~Jy is the 90\% completeness limit calculated near the ONC by \citet{stu15}. Although the completeness limit is less stringent in other regions, all sources fainter than this limit are excluded from our analysis to permit region-to-region comparison. \label{f.flux_correlate}}
\end{figure}

Figure~\ref{f.flux_correlate} plots the 70 \micron\ flux density against the 160~\micron\ to 70 \micron\ flux density ratio for protostars in our sample that were detected at both wavelengths. By attempting to detect artificial sources in the 70 \micron\ images, \citet{stu15} estimated that in their regions 1 and 2 (our OMC 2/3 and ONC~South), more than 90\% of protostars brighter than 0.12 Jy at 70 \micron\ were detected. Elsewhere in Orion A, the limit was lower, 0.03~Jy, as expected from the reduced nebulosity. Limits have not been calculated for Orion B, but we do not expect them to be more severe than 0.12~Jy, because it has low nebulosity compared to the ONC. (The NGC 2024 \ion{H}{2} region that covers a small fraction of the NGC 2023/24 region's area may be an exception.) In the following analysis, we ignore all sources fainter than 0.12~Jy at 70 \micron\ to reduce the impact of region-dependent completeness on the results.

\begin{figure}
\includegraphics[width=\hsize]{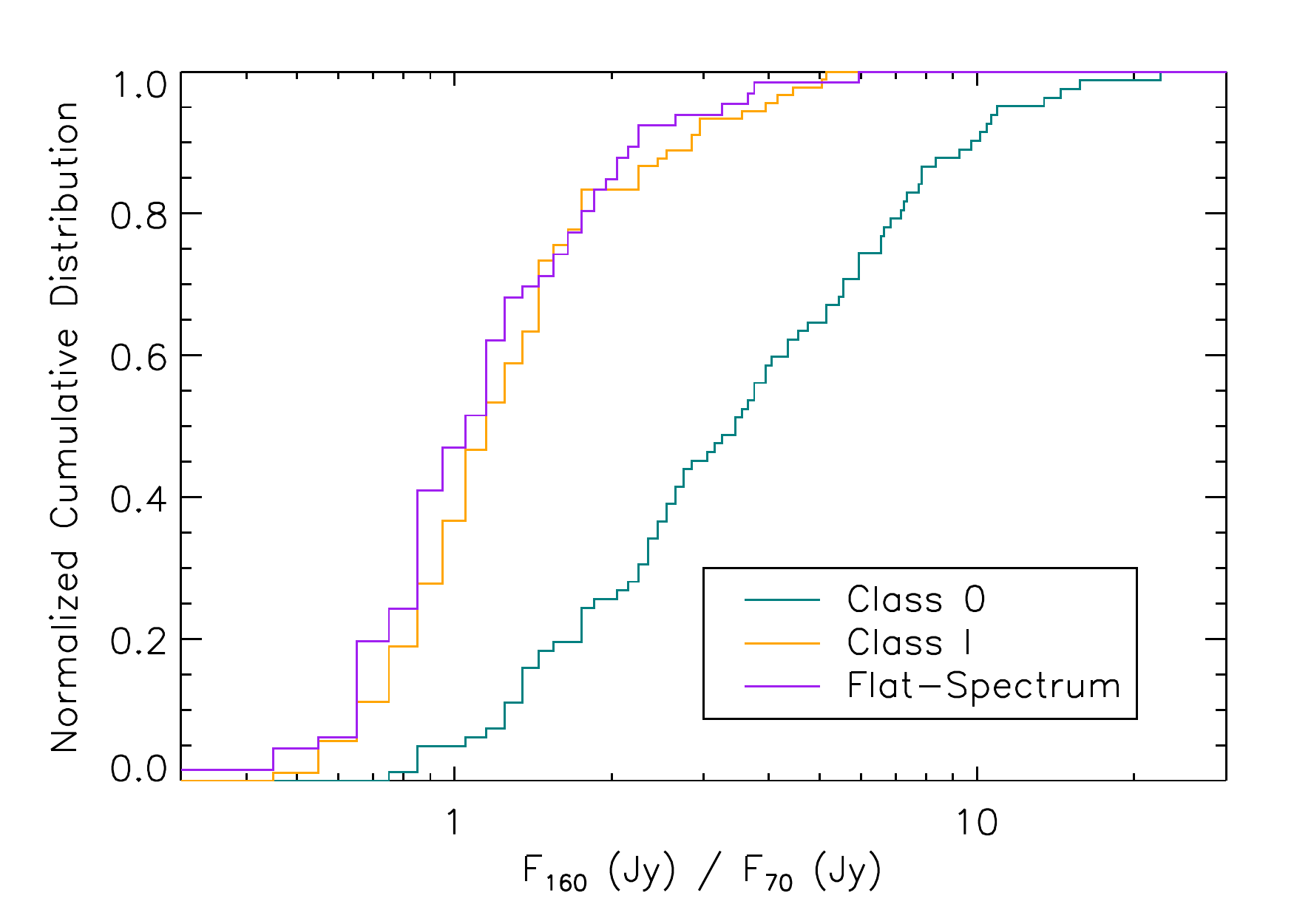}
\caption{Cumulative distributions of the ratio of 160 \micron\ to 70 \micron\ flux density for protostars in each evolutionary class. We require a flux density in excess of 0.12 Jy at 70~\micron\ and a detection at 160 \micron, so the plot is based on 82 of 92 Class~0 protostars, 90 of 125 Class~I protostars, and 66 of 102 flat-spectrum protostars. The Class 0 protostars are significantly redder in the far IR than protostars of the other two classes.\label{f.ratio_by_class}}
\end{figure}

In Figure~\ref{f.flux_correlate}, Class 0 protostars occupy a distinct region of the space and are significantly redder than other protostars when the flux density at 70 \micron\ is $\lesssim 10$ Jy.  The decreasing color with increasing flux was predicted by synthetic photometry of a grid of protostar models in \citet{ali10}, where, holding other parameters such as envelope density and outflow cavity opening angle constant, increasing the luminosities led to smaller 160~\micron\ to 70 \micron\ flux density ratios. The figure also shows that the flux density ratios of Class I and flat-spectrum protostars are indistinguishable and lack a strong dependence on flux density.

The cumulative distribution functions in Figure~\ref{f.ratio_by_class} and associated Kolmogorov-Smirnov (KS) tests confirm these class differences. The probabilities that the Class~0 flux density ratios are drawn from the same underlying distribution as the Class I and flat-spectrum flux density ratios are $1\times10^{-14}$ and $1\times10^{-13}$, respectively. Because the probabilities are very low, we conclude that the Class~0 protostars have significantly different flux density ratios from the other classes. The KS probability that the Class I and flat-spectrum flux density ratios are drawn from the same distribution is 0.17, so these classes are indistinguishable. The general trend is not surprising, because classification by evolutionary state is based on the shape of the SED. With $T_{\rm bol}<70$ K, Class~0 SEDs peak at a range of wavelengths in the far-IR and can have a range of flux density ratios there, while the SEDs of more evolved protostars peak at shorter wavelengths and cover a narrower range of smaller far-IR flux density ratios. More remarkable is the large separation between Class 0 protostars and other protostars for this flux density ratio alone.

\begin{figure}
\includegraphics[width=\hsize]{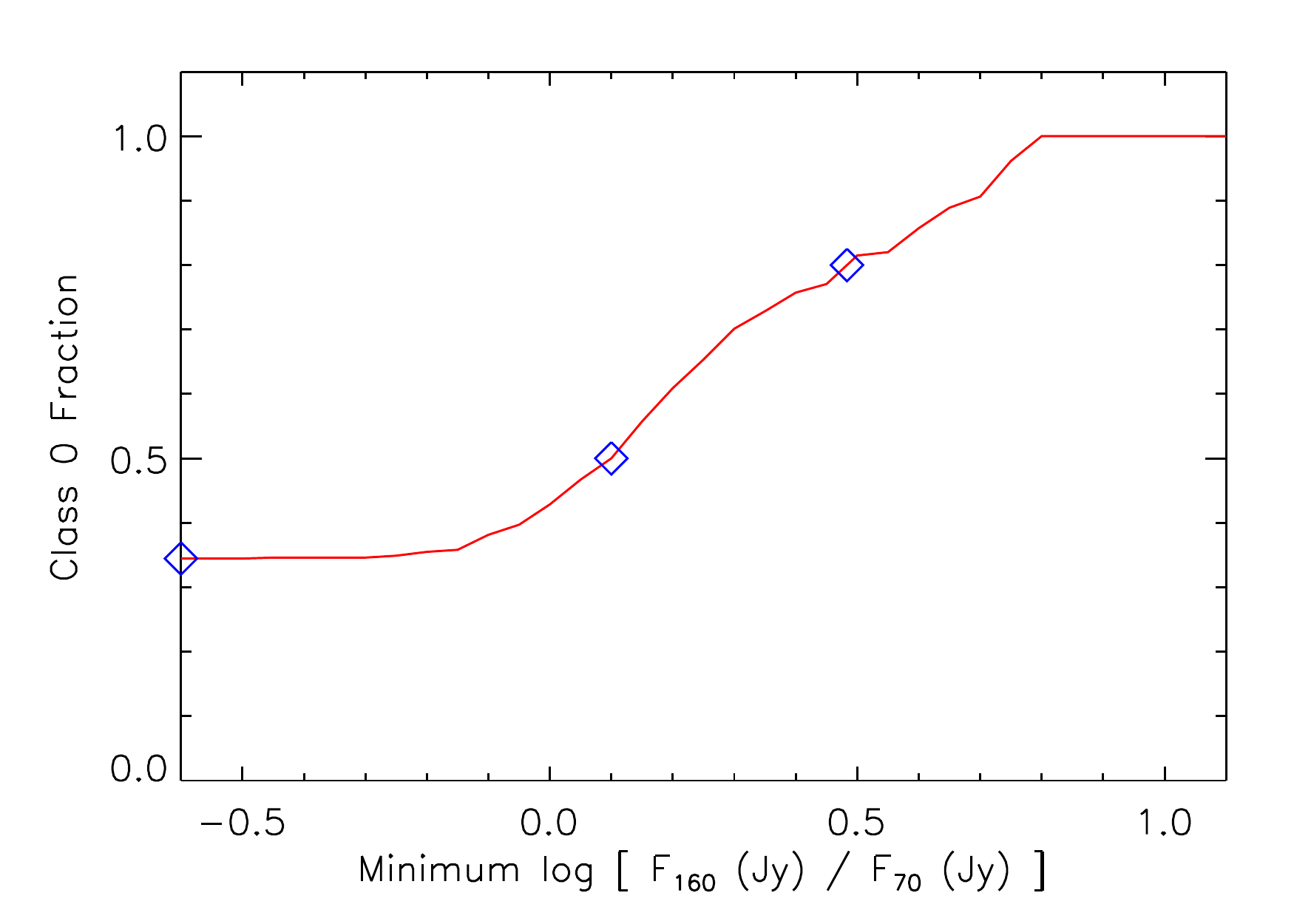}
\caption{Fraction of protostars that are Class 0 as a function of minimum flux density ratio. When considering all 238 protostars detected in both bands that satisfy the completeness limit (see text), 34\% of protostars are Class~0 (left symbol). When limiting the sample to include only those with larger flux density ratios (redder colors), an increasing fraction of protostars are Class 0. Based on the ratio of these two flux densities alone, 50\% of a large sample of protostars with $\log\ [F_{160} / F_{70}]>0.10$ (in $F_\nu$ units; middle symbol) and 80\% of such a sample with $\log\ [F_{160} / F_{70}]>0.48$ (right symbol) are likely to be Class 0.\label{f.color_cut}}
\end{figure}

This separation highlights the value of far-IR data for identifying the youngest, most rapidly accreting protostars. It circumvents the need to obtain data with broad wavelength coverage from multiple telescopes to calculate the usual diagnostics of bolometric temperature or submillimeter to bolometric luminosity ratio. Furthermore, the far-IR ratio is relatively insensitive to foreground extinction and the inclination of the protostar \citep{ali10,stu15}, factors that strongly affect the overall shape of the SED \citep[e.g.,][]{whi03,fur16}. 

Figure~\ref{f.color_cut} shows how the fraction of Class 0 protostars is higher in samples with larger 160 \micron\ to 70 \micron\ flux density ratios. When all protostars detected in both bands are included, the Class~0 fraction is 34\%, similar to what has been found in other studies of different star-forming regions \citep{dun14}, but slightly larger because we have excluded 160 \micron\ non-detections. In a sample of protostars with $\log\ [F_{160} / F_{70}]>0.10$ (in $F_\nu$ units), however, 50\% are likely to be Class~0. When $\log\ [F_{160} / F_{70}]>0.48$, 80\% are likely to be Class~0.

\subsection{Trends with Location and Environment}\label{s.loc}

\begin{figure}
\includegraphics[width=\hsize]{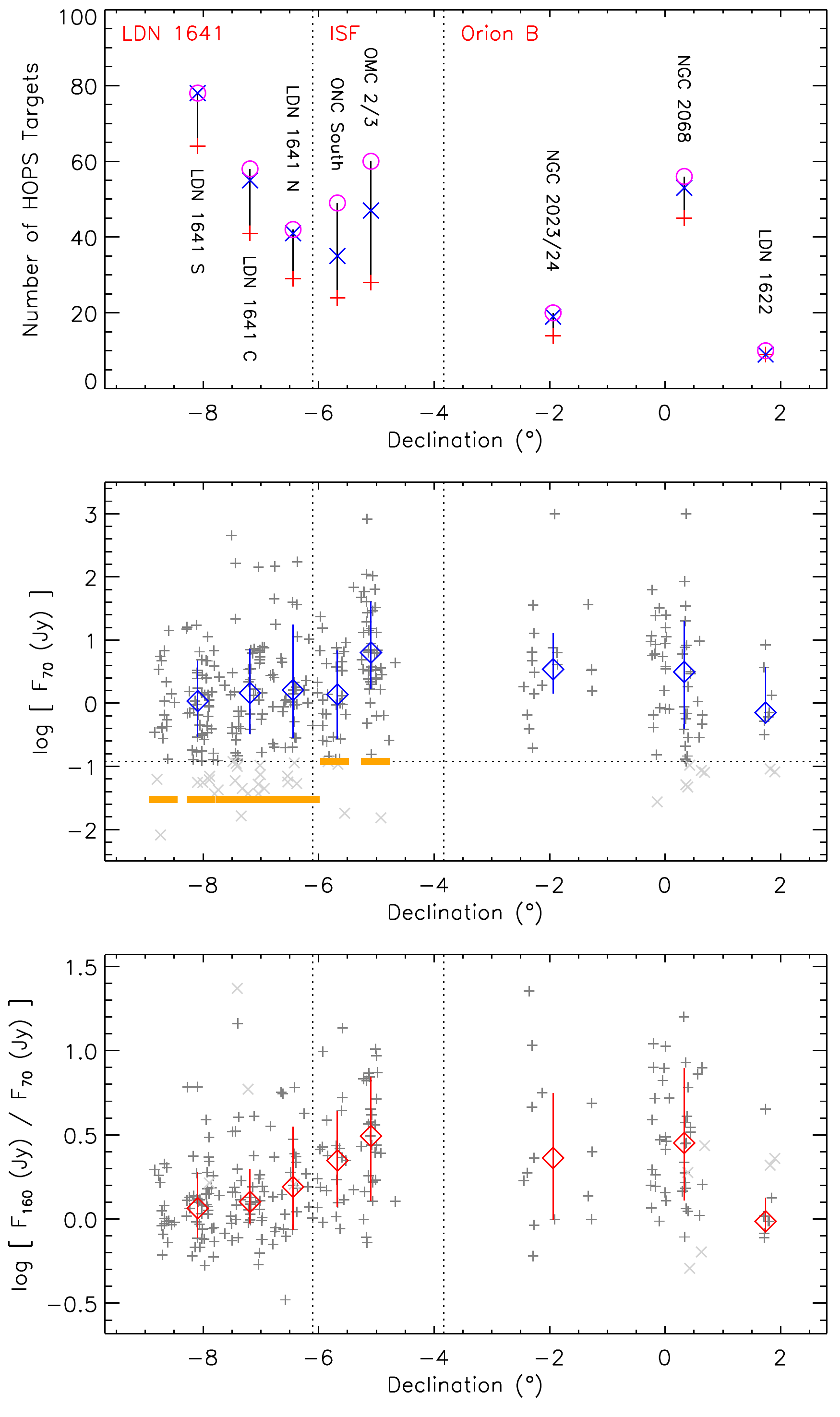}
\caption{Properties of the HOPS targets plotted by declination. Sources are binned by the regions listed in Table~\ref{t.regions}. {\em Top:} Number of targets per region. Magenta indicates all targets that were observed, blue indicates those with 70 \micron\ detections, and red indicates those with 160 \micron\ detections (which are also 70 \micron\ detections in 253 of 254 cases). {\em Middle:} Flux densities at 70 \micron. The thick orange lines show the completeness limits determined by \citet{stu15}. Sources below the dotted line are ignored in the analysis to reduce the effect of varying completeness. {\em Bottom:} Ratios of 160 \micron\ to 70 \micron\ flux densities. Sources that were excluded from the analysis in the middle panel are marked with $\times$ symbols and also excluded in the calculation of statistics shown here. In the bottom two panels, diamonds mark the median in each region, and vertical lines extend over the middle 68\% of the data in each region.\label{f.props_by_decl}}
\end{figure}

The top panel of Figure~\ref{f.props_by_decl} shows how the numbers of HOPS targets detected at 70 \micron\ and 160~\micron\ depend on their location within Orion. (In this figure, we consider all targets, including those that are less likely to be protostars.) Declination is plotted as a proxy for region; see Table~\ref{t.regions} for the declinations covered by each region. The number of targets observed per region varies between ten in LDN~1622 and 78 in LDN 1641 S, while the fraction detected at 70 \micron\ ranges from 71\% in ONC South to 100\% in LDN~1641~S. The fraction detected at 160~\micron\ ranges from 47\% in OMC 2/3 to 90\% in LDN 1622. At both wavelengths, the detection fractions are much lower in the two regions on the periphery of the ONC (ONC~South and OMC 2/3), consistent with the finding of \citet{stu15} that brighter nebulosity makes far-IR detections of protostars more challenging.

The lower two panels in Figure~\ref{f.props_by_decl} show the distributions of 70 \micron\ flux densities and 160 \micron\ to 70 \micron\ flux density ratios for each region. The middle panel shows the effect on the sample of the completeness cut discussed above. Of the 337 detections at 70 \micron, \citet{fur16} classified 319 as protostars, so the distribution of HOPS protostars is similar to what is plotted. Of the 254 detections at 160~\micron, 253 are also detected at 70~\micron.

In Orion A (the five southernmost regions in Figure~\ref{f.props_by_decl}), the brightest and reddest protostars are in OMC 2/3. Both the 70~\micron\ flux densities and the 160 \micron\ to 70~\micron\ flux density ratios decrease from north to south. In Orion B, the flux density ratios and flux densities for NGC 2023/24 and NGC~2068 are similar to those in the northernmost reaches of Orion A. The flux density ratios and flux densities for LDN 1622 are among the lowest observed, although analysis in this region is subject to a severely limited sample size.

For better statistics in the remaining analysis, we combine the regions listed in Table~\ref{t.regions} into three super-regions. LDN 1622, NGC 2028, and NGC 2023/24 are combined into ``Orion B.'' OMC 2/3 and ONC South are combined into ``ISF.'' All of the LDN 1641 sub-regions are combined. We also focus strictly on the sources classified as protostars.

\begin{figure}
\includegraphics[width=\hsize]{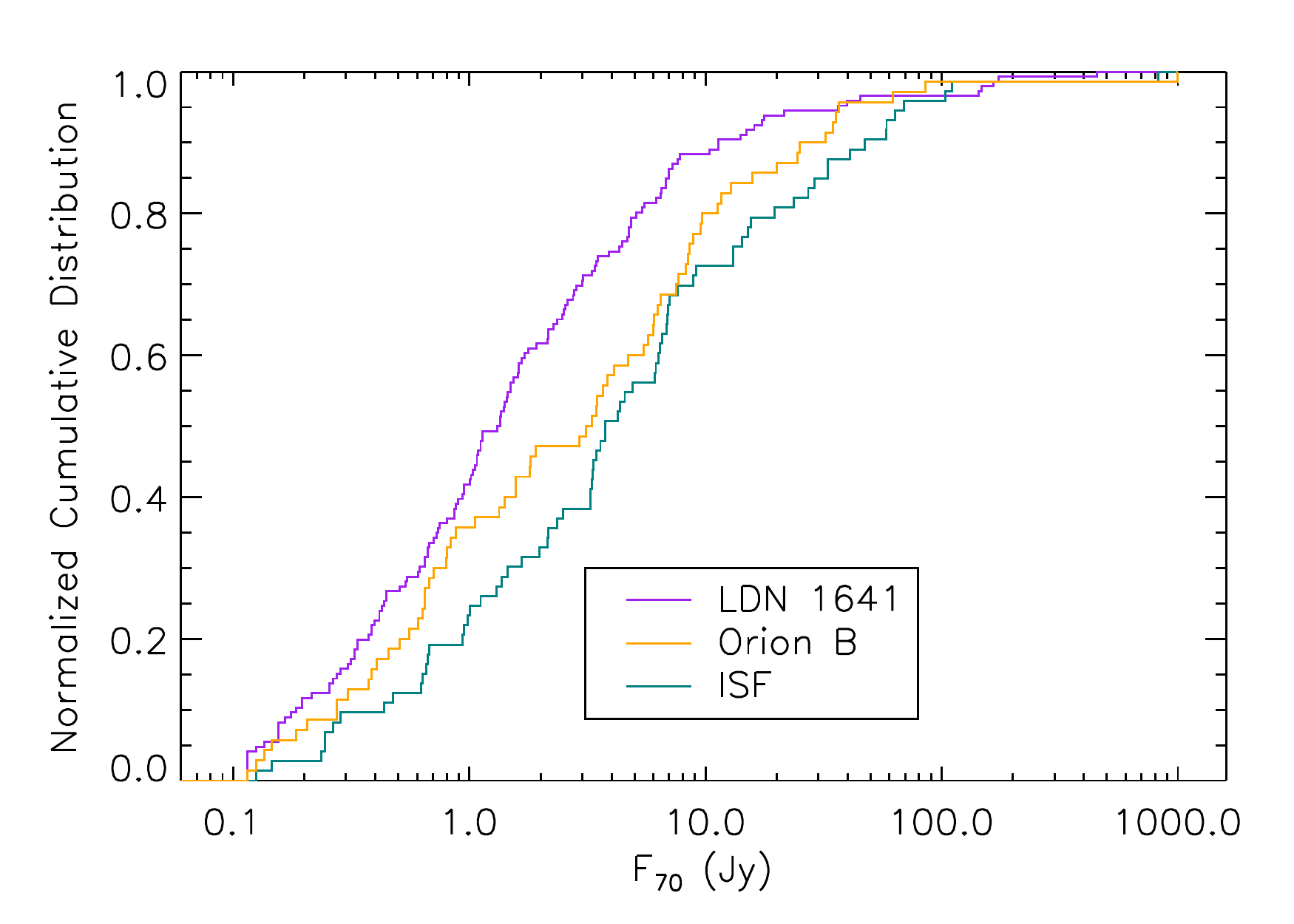}
\caption{Cumulative distributions of the 70 \micron\ flux density for protostars in each super-region. We require a flux density in excess of 0.12 Jy at 70 \micron, so the plot is based on 146 of 165 protostars in LDN 1641, 71 of 78 protostars in Orion B, and 73 of 76 protostars in the ISF. The ISF protostars are marginally brighter at 70 \micron\ than the Orion B protostars, which are in turn brighter than the LDN 1641 protostars.\label{f.flux_by_region}}
\end{figure}

\begin{figure}
\includegraphics[width=\hsize]{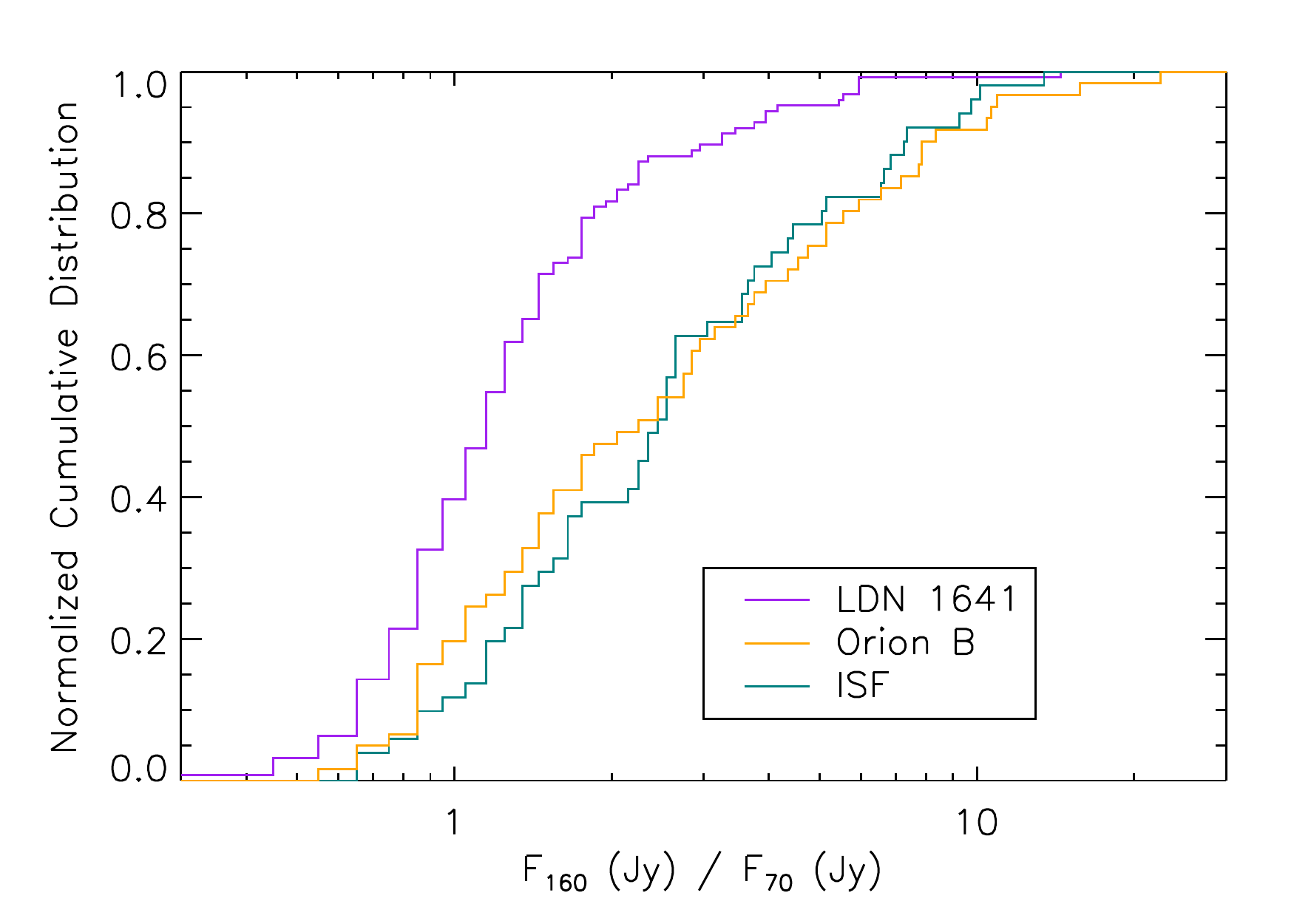}
\caption{Cumulative distributions of the ratio of 160 \micron\ to 70 \micron\ flux density for protostars in each super-region. We require a flux density in excess of 0.12 Jy at 70 \micron\ and a detection at 160 \micron, so the plot is based on 126 of 165 protostars in LDN 1641, 61 of 78 protostars in Orion B, and 51 of 76 protostars in the ISF. The ISF and Orion B protostars have similar colors and are redder than the LDN 1641 protostars.\label{f.ratio_by_region}}
\end{figure}

Figure~\ref{f.flux_by_region} shows cumulative distributions of the 70 \micron\ flux density for protostars in each super-region. It shows that the ISF protostars are marginally brighter at 70 \micron\ than the Orion B protostars, which are in turn brighter than the LDN 1641 protostars. The KS probabilities that the ISF flux densities are drawn from the same underlying distribution as the Orion B and LDN 1641 flux densities are 0.29 and $3.5\times10^{-5}$, respectively. The KS probability that the Orion B and LDN 1641 flux densities come from the same distribution is $8.8\times10^{-3}$.

Figure~\ref{f.ratio_by_region} shows cumulative distributions of the ratio of 160 \micron\ to 70 \micron\ flux density for protostars in each super-region. It shows that the ISF and Orion B protostars are redder than the LDN 1641 protostars. The KS probability that the ISF and Orion B flux density ratios are drawn from the same underlying distribution is 0.72. The KS probabilities that the LDN 1641 ratios are drawn from the same underlying distribution as the ISF and Orion B ratios are $4.4\times10^{-7}$ and $1.3\times10^{-5}$, respectively. These trends clearly demonstrate systematic variations in the properties of protostars across the Orion clouds. Most significantly, the far-IR colors of the protostars are redder for the ISF and Orion B regions than for the LDN 1641 region. Furthermore, the 70 \micron\ flux distributions for all three clouds are different, with the ISF containing the brightest protostars, then Orion~B, then LDN 1641. Of particular note is the systematic variation of the protostellar properties across the Orion~A cloud.

\section{Discussion}\label{s.disc}

The parameter most likely to be responsible for the systematic variation in colors is envelope density.  This can be established by reviewing the findings of \citet{ali10}, who used a grid of radiative transfer models (later expanded and used to fit the SEDs of the HOPS protostars by \citealt{fur16}) to examine how protostellar properties affect the \Herschel\ colors. Important properties that determine the colors are the inclination of the outflow cavity, the opening angle of the outflow cavity, the luminosity of the protostar, and the density of the envelope (usually stated at 1 AU or 1000 AU based on models that assume a radius$^{-3/2}$ density law; see \citealt{fur16}).

Examining these four properties, systematic variations in inclination are unlikely, since we expect the protostars to be randomly oriented. Outflow cavity sizes may vary systematically with region; however, the cavity opening angle has a relatively small effect on the far-IR colors. As luminosity increases, \citet{ali10} showed that the 70 \micron\ flux density increases while the 160 \micron\ to 70 \micron\ flux density ratio decreases, the opposite of the correlations seen in Figures~\ref{f.props_by_decl}, \ref{f.flux_by_region}, and \ref{f.ratio_by_region}. This suggests that the changing colors trace systematic changes in the distribution of envelope densities across the Orion clouds.

Within Orion A, flux densities and flux density ratios increase from south to north, showing an environmental dependence, with location as a proxy for environment. Similarly, \citet{stu15} found a systematic variation in the ratio of Class~0 to Class I protostars across the Orion A cloud, with the highest ratio toward the OMC 2/3 and ONC South regions. \citet{stu15} and \citet{stu16} demonstrated that, along with a larger Class 0 fraction, the ISF has higher column densities than LDN 1641. \citet{stu16} and \citet{gon19} further showed that the ISF has both higher densities and distinct gas kinematics compared to LDN 1641. \citet{car00}, \citet{meg16}, and \citet{stu18} showed that the ONC has the highest stellar densities in Orion A. Furthermore, \citet{pok20} found that the gas and stellar densities in Orion~A are strongly correlated.

The Orion B protostars are similar to those in the ISF in the distribution of their far-IR colors. Orion B, specifically NGC 2068, contains an excess of \Herschel-detected protostars that are too deeply embedded to have been detected with \Spitzer\ \citep{stu13} and have morphological evidence of youth \citep{kar20}. Like the ISF, Orion B is dominated by young stellar clusters, in contrast to the young stars in LDN 1641, which are primarily found in smaller groups or in relative isolation \citep{meg16}. Also like the ISF, the Orion B cloud may be directly influenced by the massive stars in the Orion OB1 association (e.g., \citealt{bro94}).

One explanation for the systematic changes from region to region in the typical envelope densities is that the star formation rate (SFR) is varying with time in different ways across the Orion clouds. Regions with a rising SFR will show a higher ratio of younger (Class~0) to older (Class I) protostars and, therefore, higher median envelope densities than regions with a falling SFR. For example, if the Class 0 lifetime is 0.15~Myr \citep{dun14}, a region with a constant SFR would have a Class 0 fraction of 30\% at 0.5~Myr (see Fig.\ 3 of \citealt{stu15}). In contrast, a region with a star formation episode that started 0.5 Myr ago and an SFR that increased linearly with time would have a Class 0 fraction of 51\%. The exact fraction is determined by the time since the onset of star formation and how the SFR varies with time. Given the unusually low star formation efficiency in Orion B \citep{meg16} and the high fraction of very young protostars there \citep{stu16,kar20}, the Orion B cloud may be undergoing a rapid rise in the SFR. The OMC 2/3 region in the ISF, which is rich in young protostars, may also be undergoing such a rise.

Alternatively, regions with overall higher gas density may have systematically more massive envelopes that have shorter free-fall times and collapse to form stars more rapidly \citep{kry12,dun14}. This inference is consistent with the distributions of bolometric temperatures and envelope masses for these different regions as found in an analysis of the HOPS protostars by \citet{fis17}. In contrast to the previous scenario, the SFR in this case is similar from region to region, but the resulting protostellar envelopes are more massive in regions with higher Class~0 fractions. Since the ISF region is known to have the highest gas densities in the Orion~A cloud \citep{wil99,stu15,stu16,hac20}, this could explain the high median envelope densities in the ISF. It is also possible that both effects are operating, with, for example, the rise in the SFR resulting in the high envelope densities in Orion B and the overall high gas densities resulting in the high envelope densities in the ISF. In either case, the variations demonstrate that the properties of protostars are correlated with environmental conditions and can potentially be used as tracers of the star formation history (the first scenario) or the effect of environment on the star formation process (the second scenario).

\section{Conclusions}\label{s.conc}

We described the far-IR photometry of Orion protostars obtained as part of the \Herschel\ key program HOPS. The ratio of 160 \micron\ to 70 \micron\ flux densities is an effective means of classifying protostars. Class 0 protostars occupy a distinct region of the 70 \micron\ flux versus 160~\micron\ to 70 \micron\ flux ratio diagram and show an inverse correlation between the two quantities; i.e., fainter Class 0 protostars are redder. More evolved protostars lack such a correlation. Additionally, Class 0 protostars have significantly larger 160 \micron\ to 70~\micron\ flux density ratios than Class I and flat-spectrum protostars, which are statistically indistinguishable at these wavelengths. This shows that Class 0 protostars are fundamentally different from their more evolved counterparts. In a population of protostars, 80\% of those with $\log\ [F_{160} / F_{70}]>0.48$ (in $F_\nu$ units) are likely to be Class 0. This finding circumvents the need for large multiwavelength datasets to reliably identify the most embedded protostars.

We found that redder and brighter protostars are preferentially located in the ISF and the two southern regions of Orion B relative to LDN 1641. This is consistent with the work of others, who show larger Class 0 / Class~I fractions in those regions as well as greater gas and stellar densities. These traits can be interpreted as evidence for increasing star formation rates in the ISF and Orion B or as a tendency for more massive envelopes to be inherited from denser birth environments.

\acknowledgments

We thank Babar Ali for his substantial contributions in designing the PACS survey, carrying out the data processing and the photometry, and writing the initial version of this paper while he was a staff scientist at the NASA Herschel Science Center.

Support of WJF and STM for this work was provided by NASA through awards issued by the Jet Propulsion Laboratory, California Institute of Technology. AMS acknowledges funding through Fondecyt regular (project code 1180350) and Chilean Centro de Excelencia en Astrof\'isica y Tecnolog\'ias Afines (CATA) BASAL grant AFB-170002. MO acknowledges financial support from the State Agency for Research of the Spanish MCIU through the AYA2017-84390-C2-1-R grant (co-funded by FEDER) and through the Center of Excellence Severo Ochoa award for the Instituto de Astrofísica de Andaluc\'ia (SEV/2017/0709). The National Radio Astronomy Observatory is a facility of the National Science Foundation operated under cooperative agreement by Associated Universities, Inc.

The \Herschel\ spacecraft was designed, built, tested, and launched under a contract to the European Space Agency (ESA) managed by the \Herschel/\textit{Planck} project team by an industrial consortium under the overall responsibility of the prime contractor Thales Alenia Space (Cannes), and including Astrium (Friedrichshafen), responsible for the payload module and for system testing at spacecraft level, Thales Alenia Space (Turin), responsible for the service module, and Astrium (Toulouse), responsible for the telescope, with in excess of a hundred subcontractors.

PACS has been developed by a consortium of institutes led by MPE (Germany) and including UVIE (Austria); KU Leuven, CSL, IMEC (Belgium); CEA, LAM (France); MPIA (Germany); INAF-IFSI/OAA/OAP/OAT, LENS, SISSA (Italy); and IAC (Spain). This development has been supported by the funding agencies BMVIT (Austria), ESA-PRODEX (Belgium), CEA/CNES (France), DLR (Germany), ASI/INAF (Italy), and CICYT/MCYT (Spain). HIPE is a joint development by the \Herschel\ Science Ground Segment Consortium, consisting of ESA, the NASA \Herschel\ Science Center, and the HIFI, PACS, and SPIRE consortia.

\facilities{\Herschel\ (PACS)}

\software{ATV \citep{bar01}, IDL Astronomy User's Library \citep{lan93}, StarFinder \citep{dio00}, HIPE \citep{ott10}}

\appendix

\section{The HOPS Catalog}\label{s.protostars}

The list of HOPS targets appears in Table~\ref{t:obs}. For completeness, we include all 410 objects with HOPS identifiers, whether or not these were actually observed with \Herschel\ or classified as protostars. \citet{fur16} identify the 330 HOPS sources that have high probabilities of being YSOs, including 319 that are likely Class~0, Class I, or flat-spectrum protostars.

In Table~\ref{t:obs}, column~1 lists the HOPS number, and columns~2 and 3 give coordinates. Column~4 identifies the region within the Orion~A and B clouds to which the HOPS source belongs, defined in Table~\ref{t.regions}. Column~5 lists the date on which the group containing the source was observed. Column~6 gives the \Herschel\ ObsIDs, two per group as described in Section~\ref{s.map}, and column~7 lists the group number. Since sources can appear in more than one group, we list the one for which the source position had the longest exposure time. (Sources mapped in ObsIDs 1342205228--29 were also mapped with shorter exposure times in 1342205230--31, designed to mitigate saturation in OMC 2/3.) Columns~8 through 10 pertain to the 70~\micron\ photometry, giving the flux, uncertainty, and photometric technique. The techniques are discussed in Appendix~\ref{s.phot}; entries indicate whether the measurement is from aperture photometry (A; Appendix~\ref{s.aper}) or PSF photometry (P; Appendix~\ref{s.psf}) or the source was not detected (X). Upper limits are given for some sources that were not detected. Columns~11 through 13 show the same information for the 160~\micron\ photometry. Sources with no ObsIDs were not covered by the maps or are duplicates of other sources.

\input{observation_table.tex}

\section{Data Processing and Map Generation}

From the \Herschel\ Science Archive,\footnote{\url{http://archives.esac.esa.int/hsa/whsa/}} we obtained data processed to Level~1. These are calibrated timelines, or readouts from individual PACS bolometers ordered sequentially by time of observation. The processing steps leading up to Level~1 are described in \citet{pog10} and in the data-processing guides for the \Herschel\ Interactive Processing System (HIPE; \citealt{ott10}). For our observations, PACS was used with a fixed readout frequency of 10 Hz. All instrumental effects had already been removed except for the low-frequency noise component (1/$f$ noise), which modifies the signal timelines by adding a drift component with an amplitude that is a power-law function of its Fourier frequency.

Our subsequent map generation mitigated the 1/$f$ noise, combined the two orthogonal scan directions for each group, and projected the timelines onto the final images of each field. All data discussed here are based on the FM7 version of the PACS calibration \citep{bal14} and processed with version 9 of the HIPE software. Our final maps have spatial scales of 1.6\arcsec\ pixel$^{-1}$ and 3.2\arcsec\ pixel$^{-1}$ for the 70~\micron\ and 160~\micron\ PACS filters, respectively. We used two approaches for map generation that are described below. They are based in part on the expectation that protostars will be unresolved by PACS at 70 and 160 \micron.

\subsection{The High-Pass Filter Technique}

The high-pass filter (HPF) technique is described by \citet{pop12} and implemented in HIPE. It blocks all temporal frequencies lower than a chosen filter width. This removes the low-frequency signal due to 1/$f$ noise as well as extended emission from astrophysical sources, but it preserves point sources with temporal frequencies higher than the chosen filter width, facilitating their photometry. For each readout in the timeline that was not flagged as a glitch or otherwise identified as problematic, the median value within a user-selected window surrounding the readout was subtracted from the signal value. We used HPF window widths of 15 readouts (1.5 s) and 20 readouts (2 s) for the 70~\micron\ and 160~\micron\ channels, respectively. Because point sources may elevate the median value in the HPF filter, we masked and excluded them from the calculation. For the aperture photometry discussed in Appendix~\ref{s.aper}, we used the HPF maps.

\subsection{The Scanamorphos Technique}

Scanamorphos is a map-making technique developed and described by \citet{rou13}. It removes the low-frequency noise by making use of the redundancy built into the observations. \citet{stu13} also used Scanamorphos maps for their study of the youngest protostars in Orion. Unlike the HPF technique, Scanamorphos preserves astrophysical emission on all spatial scales, ranging from point sources to extended structures with scales just below the map size. Scanamorphos maps are thus suitable for the analysis of both spatially extended and point sources, and we used them for the point spread function (PSF) fitting photometry discussed in Appendix~\ref{s.psf}. Figure~\ref{f.techniques} compares the 70~\micron\ maps of Group 019 generated with each technique. \citet{rou13} provides additional comparisons of the two techniques.

\begin{figure}
\includegraphics[width=\hsize]{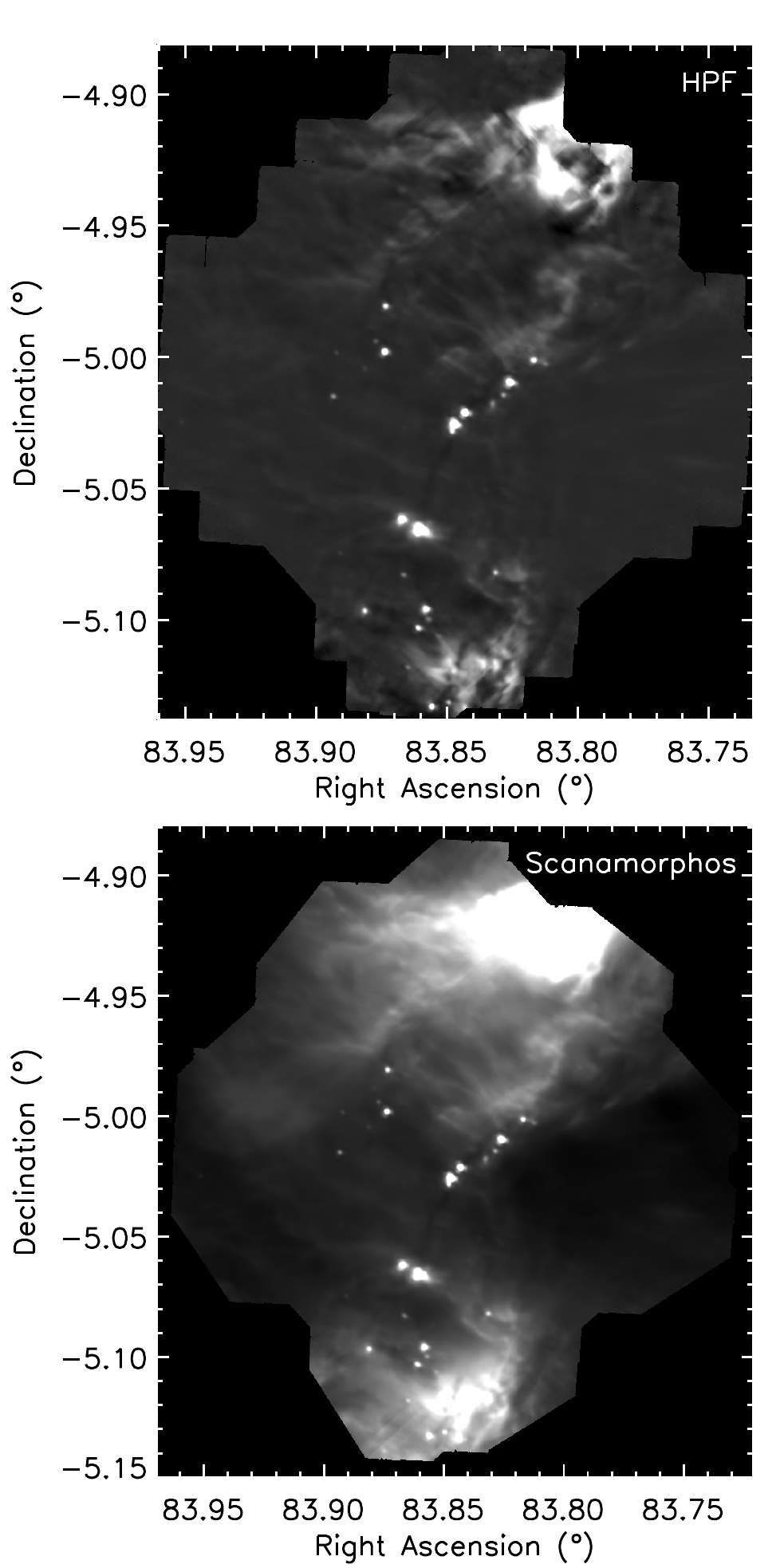}
\caption{A comparison of 70 \micron\ maps of Group 019 in OMC 2/3, generated with the high-pass filter technique ({\em top}) and the Scanamorphos technique ({\em bottom}). The high-pass filter technique preserves point sources while suppressing extended structure, and the Scanamorphos technique preserves extended structure.\label{f.techniques}}
\end{figure}

\section{Photometry}\label{s.phot}

Photometry is challenging in star-forming regions. It is necessary to distinguish spatially unresolved knots in the nebular emission from actual point sources and to estimate the local background contribution. The nebular emission is usually complex and contains gradients of emission at all spatial scales that violate the common assumption of constant or smoothly varying backgrounds. It is also difficult to disentangle contributions from multiples with separations similar to the angular resolution of the image. We resolved these issues with multiple approaches to map-making and photometry. The photometry catalog was created by selecting the best approach for each source.

\subsection{Aperture Photometry}\label{s.aper}

We used the HPF images for aperture photometry. With ATV \citep{bar01}, we manually determined the appropriate parameters for the process, and with PhotVis \citep{gut08}, we automatically identified sources and performed the photometry. Both procedures are based on {\tt aper.pro} from the IDL Astronomy User's Library \citep{lan93}. To avoid nebular contamination from the local environment, we used annuli that are small and adjacent to the source to estimate the background contribution. This step necessitated customized aperture corrections, since the background annuli include a significant fraction of a source's PSF profile. We used a calibration image of the asteroid Vesta (PACS's PSF standard; see \citealt{lut15}) to calculate aperture corrections.

We set the inner radius of the sky annulus to the aperture radius to ensure the sky annulus samples the spatially varying nebulosity near the source. At 70~\micron, we used an aperture radius of 9.6\arcsec, with the sky annulus extending from 9.6\arcsec\ to 19.2\arcsec. At 160~\micron, we used an aperture radius of 12.8\arcsec, with the sky annulus extending from 12.8\arcsec\ to 25.6\arcsec. The aperture corrections are 0.7331 and 0.6602 at 70 \micron\ and 160 \micron; the measured fluxes were divided by these factors to account for flux outside the aperture.

Aperture photometry has several benefits. It is a reliable and well-understood technique, and it is the technique used for PACS flux calibration \citep{bal14}. Further, the use of narrow apertures as described above reduces the effect of contamination and crowding for most sources. However, complex structure in the images may still result in large uncertainties in the aperture photometry of a significant fraction of the sources. The primary problem is flux contamination from a neighboring source or extended emission in either the aperture or the sky annulus. Since the annuli are narrow and few pixels are available for sky estimation, this region can be particularly affected by the presence of a strong source. Contamination has a larger effect on the 160~\micron\ measurements, because there is more extended emission at that wavelength. For the strongly affected sources in crowded or confused fields, we used PSF-fitting photometry as discussed in the next subsection.

\subsection{Point Spread Function Fitting Photometry}\label{s.psf}

In PSF-fitting photometry, a known spatial profile for point sources is fit to the measured point-source profiles, and the source brightnesses are determined from the scaling needed to make the known and measured profiles agree \citep{ste87}. The primary advantage of this technique is that it disentangles the spatial profile of the source from other point sources and emission from the local environment.

This technique requires the source PSF profiles to be well characterized. The PACS PSF is highly non-axisymmetric \citep{lut15}, requiring an accurate model. Given that no entirely contamination-free sources are available in our HOPS fields, we used images of Vesta as a proxy for the PSF profiles of our sources. To remove any systematic photometry offsets between PSF-based photometry and aperture-based photometry, we repeated the PSF measurements on a subset of PACS flux standard stars. This comparison allowed us to calibrate measured PSF amplitudes with actual flux measurements.

We used the StarFinder package \citep{dio00} for fitting PSF profiles and measuring photometry of sources in the Scanamorphos-reduced images. If there were overlapping sources, we fit them simultaneously even if not all were explicit HOPS targets. The StarFinder source-finding algorithm suffers from the same challenges as other algorithms, in that many of the detected sources are partially resolved or unresolved compact structures in the nebular emission. On the other hand, the PhotVis source finder used for aperture photometry circumvents this problem by using the local confusion noise surrounding each source to set a threshold for detection. We therefore limited our PSF-fitted photometry to protostars for which aperture photometry was known to be unreliable.

\subsection{Final Photometry and Uncertainties}\label{s.final}

Since the aperture photometry might be contaminated by the presence of other nearby sources and spatially extended nebular emission, we inspected the images of each HOPS protostar individually. This inspection identified contaminants such as other point sources within the aperture radius or strong nebular features that were likely to affect photometry. Once identified as contaminated, we considered both PSF-fitted and aperture photometry in the context of the overall SED for the source. Aperture photometry for contaminated sources was rejected in favor of PSF-fitted photometry.

Of the 373 unique HOPS targets observed by PACS, we detected 338 (91\%) in at least one of the 70 \micron\ and 160 \micron\ bands. Of the 338 detections, 253 (75\%) were detected in both bands, 84 (25\%) were detected at only 70 \micron, and one was detected at only 160 \micron. Of the 337 detections at 70 \micron, 276 were measured with aperture photomery, and 61 were measured with PSF-fitting photometry. Of the 254 detections at 160 \micron, 150 were measured with aperture photomery, and 104 were measured with PSF-fitting photometry. The expanded role of PSF-fitting photometry at 160 \micron\ reflects how the colder protostellar envelope dust probed at longer wavelengths is less distinct from the dense cloud structures in which the protostars are forming. It is also a consequence of the lower resolution of the 160 \micron\ maps and, therefore, higher likelihood of source blending compared to the 70 \micron\ maps.

Aperture photometry uncertainties are calculated using the root-mean-square (RMS) variation in the intensity values within the sky annulus, but with a floor of 5\%. PSF-fitting photometry uncertainties are set to 10\% of the measured flux. When a source was observed more than once, only the observation with the longest exposure time appears in Table~\ref{t:obs}. We averaged all individual measurements to obtain the final estimate of its flux, and the reported uncertainty is the RMS of the individual uncertainties.

\end{document}

%% file: observation_table.tex
\begin{longrotatetable}
\begin{deluxetable*}{@{\extracolsep{4pt}}ccccccccccccc@{}}
\tabletypesize{\scriptsize}
\tablewidth{0pt}
\tablecaption{HOPS Sources\label{t:obs}}
\tablehead{
\colhead{} &
\colhead{R.A.} &
\colhead{Decl.} &
\colhead{} &
\colhead{Obs.\ Date} &
\colhead{} &
\colhead{} &
\multicolumn{3}{c}{70 \micron} &
\multicolumn{3}{c}{160 \micron} \\
\cline{8-10}
\cline{11-13}
\colhead{HOPS} & 
\colhead{($^\circ$)} &
\colhead{($^\circ$)} &
\colhead{Region} &
\colhead{(UT)} &
\colhead{ObsID} &
\colhead{Group} &
\colhead{Flux (Jy)} &
\colhead{Unc.\ (Jy)} &
\colhead{Type\tablenotemark{1}} &
\colhead{Flux (Jy)} &
\colhead{Unc.\ (Jy)} &
\colhead{Type\tablenotemark{1}}
}
\startdata
000                                & 88.6171 & \phantom{$-$}1.6264 & LDN 1622    & \nodata     & \nodata         & \nodata & \nodata     & \nodata     & \nodata & \nodata     & \nodata     & \nodata \\
001                                & 88.5514 & \phantom{$-$}1.7099 & LDN 1622    & 2011 Mar 06 & 1342215365$-$66 & 000     & 3.697e$+$00 & 1.850e$-$01 & A       & 3.836e$+$00 & 2.039e$-$01 & A       \\
002                                & 88.5380 & \phantom{$-$}1.7144 & LDN 1622    & 2011 Mar 06 & 1342215365$-$66 & 000     & 5.188e$-$01 & 2.617e$-$02 & A       & 4.013e$-$01 & 4.013e$-$02 & P       \\
003                                & 88.7374 & \phantom{$-$}1.7156 & LDN 1622    & 2011 Apr 18 & 1342218780$-$81 & 001     & 3.187e$-$01 & 1.622e$-$02 & A       & 2.621e$-$01 & 1.628e$-$02 & A       \\
004                                & 88.7240 & \phantom{$-$}1.7861 & LDN 1622    & 2011 Apr 18 & 1342218780$-$81 & 001     & 6.116e$-$01 & 3.083e$-$02 & A       & 5.926e$-$01 & 3.290e$-$02 & A       \\
005                                & 88.6340 & \phantom{$-$}1.8020 & LDN 1622    & 2011 Apr 16 & 1342218703$-$04 & 003     & 7.103e$-$01 & 3.573e$-$02 & A       & 6.885e$-$01 & 4.186e$-$02 & A       \\
006                                & 88.5767 & \phantom{$-$}1.8176 & LDN 1622    & 2011 Apr 16 & 1342218703$-$04 & 003     & 9.110e$-$02 & 5.523e$-$03 & A       & 1.900e$-$01 & 1.900e$-$02 & A       \\
007                                & 88.5835 & \phantom{$-$}1.8452 & LDN 1622    & 2011 Apr 16 & 1342218703$-$04 & 003     & 1.342e$+$00 & 6.728e$-$02 & A       & 1.792e$+$00 & 9.277e$-$02 & A       \\
\enddata
\tablenotetext{1}{A = aperture photometry; P = PSF-fitted photometry; X = not detected (but upper limits may have been determined).}
\tablenotetext{2}{HOPS 109, 111, 212, and 362 are duplicates of HOPS 40, 60, 211, and 169, respectively.}
\end{deluxetable*}
\end{longrotatetable}